\documentclass[aps,prl,twocolumn,showpacs]{revtex4}

\usepackage{epsfig}
\usepackage{graphicx}

\DeclareGraphicsExtensions{.eps,.EPS}

\begin{document}

\title{Radio-frequency association of molecules: the dressed state picture}
\author{Q. Beaufils, A. Crubellier$^{\diamond}$, T. Zanon, B. Laburthe-Tolra, E. Mar\'echal, L. Vernac, and O. Gorceix}
\affiliation{Laboratoire de Physique des Lasers, CNRS UMR 7538, Universit\'e Paris 13,
99 Avenue J.-B. Cl\'ement, 93430 Villetaneuse, France}
\affiliation{$^{\diamond }$ Laboratoire Aim\'e Cotton, CNRS II, B\^atiment 505, Campus d'Orsay, 91405 Orsay Cedex, France }

\begin{abstract}
We develop a theoretical model to describe the radio-frequency (rf) induced coupling of a pair of colliding atoms to a Feshbach molecule when a magnetic field arbitrarily far from the Feshbach resonance is modulated in time. We use the dressed atom picture, and show that the coupling strength in presence of rf is equal to the Feshbach coupling strength multiplied by the square of a Bessel function. The argument of this function is equal to the ratio of the atomic rf Rabi frequency to the rf frequency. We experimentally demonstrate this law by measuring the rate of rf-association of molecules using a Feshbach resonance in d-wave collisions between ultra-cold chromium atoms.  
\end{abstract}


\date{\today}

\maketitle

In a Feshbach resonance, pairs of colliding atoms are resonantly coupled to a molecular bound state, so that one can produce ultra-cold molecules, either by pulsing a magnetic field \cite{donley}, or by ramping a magnetic field through the resonance \cite{Kohler}. In this context, radio frequency (rf) oscillating magnetic fields have first been employed to dissociate Feshbach molecules and precisely measure their binding energy \cite{regal}, and then to produce molecules by associating atom pairs near a Feshbach resonance (\cite{thomson},\cite{Ospelkaus}) or to transfer ground state molecules between states of different vibrational quantum numbers \cite{lang}. Thus rf has become one of the major tools used for the production of weakly bound Feshbach molecules. These molecules can then be coherently transfered to their rovibrational ground state at high phase-space densities \cite{Ni}, with important perspectives for the study of dipolar gases \cite{dipolar}, quantum information \cite{Demille}, or metrology \cite{Ye}.

In absence of spin-dependent interactions, coupling between different molecular potentials is not provided by rf, which merely induces precession of the total spin of a molecule. Rf alone does not couple two different vibrational states of a given molecular potential either, because of their orthogonality. Rf association therefore necessarily relies on the spin-dependent part of the interatomic potential; it is thus generally considered that the proximity of a Feshbach resonance is needed for rf association to be efficient. Indeed, an increasing number of experiments using rf to associate molecules \cite{increasing} are performed very close to a Feshbach resonance. Corresponding time-dependent theoretical treatements are available \cite{hanna}. They rely on numerics, and are very sensitive to the system under study. Moreover, although it has been noted that the efficiency of rf association decreases when one gets further away from resonance \cite{zirbel}, it has up to now never been shown generically \textit{how close} one needs to be from the resonance to produce molecules, and \textit{how efficient} association is as a function of the rf power and frequency. 

In this Letter, we first derive an analytical law for the strength of the coupling of pairs of colliding atoms to a molecular bound state when a magnetic field is modulated at rf frequencies. Here we do not address the interesting many-body question of the molecular conversion fraction as a function of phase space density \cite{conversion}, but focus instead on the two-body problem. When the atomic Zeeman effect is linear, we find that rf coupling is described by a very simple law: the rf coupling strength is equal to the Feshbach coupling strength times $\left(J_1\left(\Omega / \omega\right)\right)^2$, when the rf frequency $\omega$ corresponds to the difference of energy between the open and the bound channel. Association of molecules can thus be performed arbitrarily far from the Feshbach resonance, provided the available rf Rabi frequency $\Omega$ in the experiment is on the order of $\omega$. We then provide experimental evidence for this formula, using ultra-cold trapped chromium atoms near a Feshbach resonance in d-wave collisions.

Our theoretical description of rf-association uses the dressed atom picture \cite{cohen}. The Hamiltonian describing the interaction of a pair of atoms with a linear rf field parallel to a static field can be written in the following way:

\begin{equation}
H = H_{mol} + \hbar \omega_0 S_z + \hbar \omega a^+a + \lambda S_z(a+a^+)+H_{dd}
\label{hamiltonian}
\end{equation}
$H_{mol}$ is the molecular Hamiltonian with no magnetic dipole-dipole interactions. $\hbar \omega_0 S_z$ describes the Zeeman effect, which is assumed to be linear; $\hbar \omega_0 = g_J \mu_B B$,  with $g_J$ the Land\'e factor, $\mu_B$ the Bohr magneton, and $B$ the magnetic field. $ H_{rf} = \hbar \omega a^+a + \lambda S_z(a+a^+)$ accounts for the energy of the linear rf field of angular frequency $\omega$ parallel to $z$, and its coupling to the molecules. We define the Rabi frequency by $2 \lambda \sqrt{N} = \hbar \Omega$, where the number of rf photons $N$ is assumed to be large. $H_{dd}$ is the dipole-dipole interaction Hamiltonian. The key point is that the first part of the Hamiltonian $H_{mol} + \hbar \omega_0 S_z$ commutes with  $ H_{rf}$ \cite{cohen} . In absence of $H_{dd}$  the diagonalization of $H$ is exact (irrelevant of the rf power), and the eigenstates appear as a series of manifold of dressed states:

\begin{equation}
\left|\widetilde{X,N}\right\rangle = T_X^+ \left|X,N\right\rangle 
\label{dressedstates}
\end{equation}
$X$ denotes the internal state of the molecule. $T_X= \exp\left(-\frac{M_X \lambda}{\hbar \omega}\left(a-a^+\right)\right)$ is a field translation operator, with $M_X$ the spin projection of state $X$ along the axis $z$. The  eigen-energies of the dressed states $\left|\widetilde{X,N}\right\rangle$ are $W_X=E_X + M_X \hbar \omega_0 + N \hbar \omega + \frac{M_X^2 \lambda^2}{\hbar \omega}$, where $E_X$ is the eigen-energy of state $X$ without rf or magnetic field. The term $\frac{M_X^2 \lambda^2}{\hbar \omega}$ is negligible, as $N>>1$. 

We apply this model to an isolated Feshbach resonance, between molecular states $\left|A\right\rangle= \left|\psi_{\epsilon}\right\rangle$ (the incoming channel with collision energy $\epsilon$) and $\left|B\right\rangle = \left|\psi _{bound}\right\rangle$ of respective spin projections $M_A$ and $M_B$. The Feshbach resonance (without rf) occurs at a magnetic field such that $E_A-E_B+\left(M_A-M_B\right) \hbar \omega_0 = 0$. The strength of the coupling is  $\Gamma_0 (\epsilon)=2\pi |< \psi _{bound} | H_{dd} | \psi_{\epsilon} >|^2$ \cite{mies}.

In presence of rf, additional resonances occur when eigenstates belonging to different rf manifolds cross, $i.e.$ when $E_A-E_B+\left(M_A-M_B\right)\hbar \omega_0 + \left(N_A-N_B\right) \hbar \omega =0$.  The strength of the resonance coupling between those two dressed states is given by an equation similar to the one used for $\Gamma_0 (\epsilon)$:  

\begin{eqnarray}
\Gamma_{N_A,N_B}(\epsilon)= 2 \pi \left| \left\langle \widetilde{A,N_A}\right|H_{dd}\left|\widetilde{B,N_B}\right\rangle \right|^2 \nonumber\\
= 2 \pi \left| \left\langle A\right|H_{dd}\left|B\right\rangle \left\langle N_A\right|T_AT_B^+\left|N_B\right\rangle \right|^2 \nonumber\\
= \Gamma_0 (\epsilon) \left| \left\langle N_A\right|e^{\frac{\left(M_A-M_B\right)\lambda}{\hbar \omega}(a-a^+)} \left|N_B\right\rangle \right|^2 \nonumber\\
\end{eqnarray} 

We follow a calculation generalizing the one in \cite{cohen}, which consists of expanding the exponential in series, then recognize the series expansion of Bessel functions when the photon number is very large, and we find that:

\begin{equation}
\Gamma_{N_A,N_B}(\epsilon) =\Gamma_0(\epsilon) \left( J_{N_B-N_A}\left(\frac{\left(M_A-M_B\right) \Omega}{\omega}\right) \right)^2
\label{resultanne}
\end{equation}
where $J_N$ is the $N^{th}$ Bessel function. This is the main result of this paper. When $\omega > \Gamma_0 (\epsilon)$, these Feshbach resonances between dressed states are isolated, and the rf coupling strength to the Feshbach molecule with the exchange of $N_B-N_A = M$ photons is proportional to $\left( J_{M}\left(\frac{\left(M_A-M_B\right) \Omega}{\omega}\right) \right)^2$.

The appearance of a Bessel function in eq. (\ref{resultanne}) can be understood in the following way. Let us consider the case of a transition between $M_A=-1$ and $M_B=0$ for sake of simplicity; an extension to arbitrary $M_A$ and $M_B$ is straightforward. As the Zeeman effect is linear, in presence of rf the eigenstate $\left|A\right\rangle$  is phase modulated at the rf frequency, whereas $\left|B\right\rangle$ is not. The phase time dependent factor $\exp\left(i\left(\frac{(E_A - \hbar \omega_0) t}{\hbar} - \frac{\Omega}{\omega} \sin(\omega t)\right)\right)$ can be developped as $\sum_N (-i)^N J_N \left(\frac{\Omega}{\omega}\right) \exp(i (\frac{(E_A-\hbar \omega_0) t}{\hbar} + N \omega t))$. In the interaction between $\left|A\right\rangle$ and $\left|B\right\rangle$ in presence of rf, resonances occur when $E_B =E_A - \hbar \omega_0 + N \hbar \omega$, which corresponds to a resonant coupling between $\left|A\right\rangle$ and $\left|B\right\rangle$ with the absorption of N photons. The amplitude of this coupling is therefore set by the fraction of the entrance channel wavefunction evolving at frequency $E_B / \hbar$, $i.e.$ $J_N \left(\frac{\Omega}{\omega}\right)$. The Bessel function in eq. (\ref{resultanne}) thus appear as the consequence of a phase modulation of the difference of energy between the eigenstates.

There is an analogy between our result and studies which describe electric field resonant dipole-dipole collisional energy transfer in Rydberg atoms in presence of microwave. In that case, the micro-wave assisted cross-section for collisional energy transfer is also set by Bessel functions \cite{Pillet}. 

Here, we apply this formalism to the case of the Feshbach resonance in ultra-cold chromium atom collisions studied in \cite{beaufils08}. In this resonance, first observed in \cite{werner}, the incoming channel is a $l=2$ partial wave:  $\left|A \right\rangle = \left| S=6,M_{S}=-6,l=2,m_{l}=1\right\rangle \times F_{\epsilon}(R)$ is resonantly coupled to a molecular bound state $\left|B \right\rangle = \left| S=6,M_{S}=-5,l=0\right\rangle \times F_{bound}(R)$ through dipole-dipole interactions, at a magnetic field of $B_{res} = $8.155 G. $F_{\epsilon}(R)$ and $F_{bound}(R)$ are the radial wavefunctions of respectively the incoming channel and the bound molecular state. At low temperatures, coupling of the states  $\left|A \right\rangle$ and  $\left|B \right\rangle$ is strongly inhibited because of the centrifugal barrier in the incoming channel which reduces the overlap between $F_{\epsilon}(R)$ and $F_{bound}(R)$. Given the experimental atomic density, the coupling rate is then much smaller than the collisionally limited life-time of the bound state, so that association of molecules at the Feshbach resonance simply translates into losses, with a loss rate solely determined by the Feshbach coupling strength itself \cite{beaufils08}:

\begin{equation}
\frac{\dot{n}}{n}=-\alpha \left( n \Lambda_{dB}^3\right) \Gamma_0 (\epsilon_0) exp(-\epsilon_0 / k_B T)  \equiv -K_2^0(\epsilon_0) n
\label{eqpsd}
\end{equation}
where $\Lambda_{dB} = \frac{h}{\sqrt{2 \pi m k_B T}}$ is the thermal de Broglie wavelength, $\alpha = 6 \sqrt{2}$ is a numerical factor, and $\epsilon_0 = g_J \mu_B (B-B_{res})$ is the energy of the bound molecular state relative to the dissociation limit of the incoming channel. $g_J = 2$ for chromium atoms in the $^7$S$_3$ state. $\Gamma_0 (\epsilon) \propto \epsilon^{5/2}$ describes the Feshbach coupling, and $K_2^0(\epsilon_0)$ is the associated loss parameter. Analysing losses near this Feshbach resonance is thus a good means to directly measure $\Gamma_0 (\epsilon)$.

Here, we use rf fields with a rf frequency $\omega / 2 \pi$ close to $\epsilon_0 /h$. As in \cite{beaufils08}, we analyse losses, deduce a rf dependent loss parameter $ K_2(\Omega,\omega,\epsilon_0)$ and relate it to the rf-assisted coupling parameter, using eq. (\ref{eqpsd}). As the rf-assisted coupling strength is related to the Feshbach coupling strength (eq. (\ref{resultanne})), we therefore conclude that 

\begin{equation}
K_2(\Omega,\omega,\epsilon_0) = K_2^0 (\epsilon_0 + \hbar \omega) \times \left(J_1\left(\Omega / \omega\right)\right)^2
\label{eqK3_1}
\end{equation} 
which we experimentally demonstrate below. $K_2(\Omega,\omega,\epsilon_0) \times n$ is the rate of rf association of molecules.

The experimental procedure is described in \cite{beaufils08}. We load cold ground state chromium atoms in a one-beam optical dipole trap and polarize them in the lowest energy state $\left|S=3,m_S=-3\right\rangle$. Forced evaporation is performed by  transfering power from this horizontal dipole trap to a crossed vertical beam. At the end of this process, the atom number is between 50 000 and 60 000 and the temperature is 7 $\mu$K. At this point we tune a static homogeneous magnetic field $B$ near $B_{res}$. We use rf spectroscopy to calibrate the magnetic field with an uncertainty of 2 mG.

\begin{figure}
\centering
\includegraphics[width= 2.8 in]{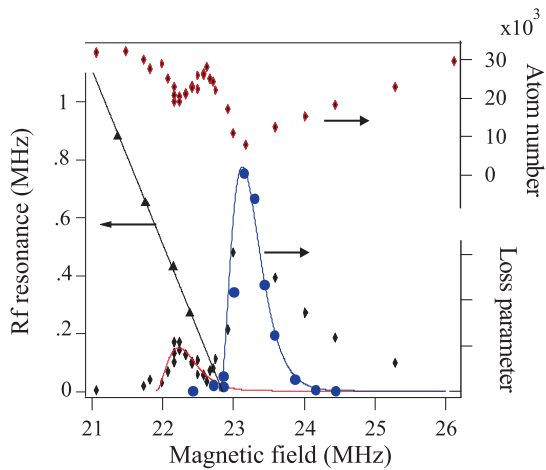}	
\caption{\setlength{\baselineskip}{6pt} {\protect\scriptsize
Losses as a function of magnetic field. We report: (up, right) the number of atoms after 8s hold time in magnetic field $B$ in presence of an rf field of 900 kHz; (down, right) black diamonds: the corresponding loss parameter; blue bullets: loss parameter without rf. The blue solid line results from a fit of the data without rf using eq. (\ref{eqpsd}). The red solid line corresponds to the blue solid line, shifted by 900 kHz, and multiplied by $\left(J_1\left(\Omega/\omega\right)\right)^2$. (left) rf spectroscopy of the Feshbach resonance: rf resonant losses vary linearly with the static magnetic field. Black solid line : linear fit of slope one.}} \label{figurelossesrf}
\end{figure}

We report in Fig \ref{figurelossesrf} the number of atoms remaining in the trap after $8$ s in presence of an rf magnetic field at $\omega/2 \pi =$ 900 kHz, as a function of $B$. When the pairs of colliding atoms are coupled to the bound molecular state, we observe losses, because the molecular state has a very short lifetime, as explained above. In this loss spectrum we observe two peaks : one due to the Feshbach resonance (which we will refer to as the 'Feshbach peak') and one shifted to lower magnetic fields (the 'rf peak'). This latter peak results from the coupling between the close and open molecular channels by the rf field, with the \textit{emission} of one rf photon. 

We also plot in Fig \ref{figurelossesrf} the corresponding loss parameter, deduced by comparing the initial number of atoms to the number of atoms remaining in the trap after 8s. The lineshape of the rf peak is very similar to the lineshape of the Feshbach resonance without rf. The separation between the two lineshapes is exactly the rf frequency. We also find that the lineshape of the loss parameter describing the 'Feshbach peak' in presence of rf is distorted as compared to when the rf is off (see Fig \ref{figurelossesrf}). In particular, more losses are observed at high magnetic fields. We do not account for these larger losses; they may be induced by the coupling between the close and open molecular channels associated with the \textit{absorption} of one or more rf photons. We have not studied this feature in detail, and the rest of the paper concentrates on the rf peak.

Tuning $B$ to a value below $B_{res}$, we measure losses as a function of the rf frequency. We observe rf-induced losses at rf frequencies depending on $B$. For a given $B$, we fit the rf loss curve using a $K_2$ of the form given in eq. (\ref{eqpsd}), and we report on Fig \ref{figurelossesrf} the fitted position of the resonance ($i.e.$ using $\epsilon_0$ as a free parameter) as a function of $B$.  We find that the resonant rf frequency varies linearly with $B$. A linear fit of slope one to the rf resonance frequency as a function of $B$ (in MHz) predicts that the 'rf peak' merges with the 'Feshbach peak' at $B=8.157$ G. This value corresponds within a very good accuracy to the previously measured position of the Feshbach resonance \cite{beaufils08}. Losses in presence of rf are thus related to population of the molecular bound state involved in the Feshbach resonance. The linear dependence versus $B$ of the relative Zeeman energy between the closed and the open channel is expected, as the value of $\Gamma_0(\epsilon)$ is typically 10 Hz at 8 $\mu K$ \cite{beaufils08}, much smaller than the typical detuning from resonance ($\omega / \Gamma_0(\epsilon)\approx 10^5$). We therefore expect to be in the regime where our theoretical model is valid.

\begin{figure}
\centering
\includegraphics[width= 2.8 in]{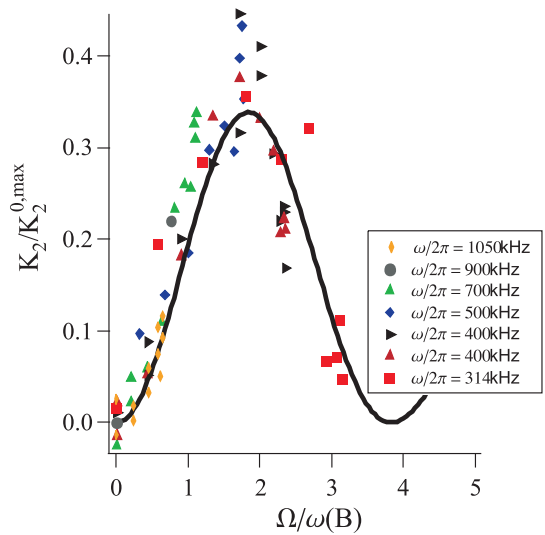}
\caption{\setlength{\baselineskip}{6pt} {\protect\scriptsize
rf-assisted loss parameter $K_2$ as a function of $\Omega / \omega (B)$ for different resonant rf frequencies, corresponding to various magnetic fields $B$. $K_2$ is scaled to the loss parameter measured at the Feshbach resonance. The solid line is $\left(J_1\left(\Omega / \omega\right)\right)^2$.}} \label{figurebessel}
\end{figure}

We now turn to the quantitative analysis of the amplitude of rf induced losses. The 'rf peak' is maximum for an rf frequency  $\omega (B)$ which varies linearly with $B$. This frequency corresponds to the maximum of the lineshape described by eq. (\ref{eqK3_1}). We find that the corresponding loss parameter $K_2^{max}$ depends both on $\Omega$ and on $\omega (B)$. We plot $K_2^{max}$  as a function of  $\Omega / \omega (B)$ in Fig \ref{figurebessel} for different resonant rf frequencies $\omega (B)$. In this figure, we have divided $K_2^{max}$  by the maximum loss parameter measured at the Feshbach resonance without rf, $K_2^{0,max} = 4 \times 10^{-20} $m$^3 $s$^{-1}$. Within the experimental error bars, this scaled loss parameter exactly fits eq. (\ref{eqK3_1}), with no adjustable parameter: $\Omega$ is measured by observing Rabi oscillations at resonance for any $\omega$.

Our framework, supported by the experimental evidence of Fig \ref{figurebessel}, treats rf association as a Feshbach resonance between dressed states. Other observations validate this treatment. In particular, we have found that all the peculiar features observed for the Feshbach resonance \cite{beaufils08} also apply to the case of the rf-resonant losses: for example, the dynamics of the rf-induced atom decay is well described assuming a two-body loss parameter; both the temperature dependence and the lineshape of the rf-resonant loss parameter also are, to within the experimental uncertainty, identical to the ones of the Feshbach resonance. This indicates that the rf association process in presence of rf is qualitatively similar to the coupling to the molecular bound state at the Feshbach resonance itself. The advantage of using rf fields to associate molecules rather than using magnetic fields sweeps near the Feshbach resonance itself simply relies on the larger versatility of rf components ($e.g.$ smaller switching and ramping times).

With sufficient rf power, one should also observe Feshbach resonances assisted by multiphoton processes. For a given magnetic field, a process with the emission of two photons happens at $\hbar \omega (B) /2 $. We indeed observed such resonances. These are unfortunately more difficult to analyse, mainly because at large rf power the rf amplifier produces non negligible power in the second harmonics, and two-photon processes at $\omega (B) /2$ coexist with one-photon processes at $2 \omega (B) /2$.

Although the agreement between our data and eq. (\ref{eqK3_1}) is good up to $\Omega / \omega \approx 4$, this agreement breaks down for higher $\Omega / \omega$. However, in practice, given our rf amplifier, $\Omega / \omega > 4 $ corresponds to $\omega < 300$ kHz, for which we observe  shifts of the molecular level, and of the Feshbach resonance, as a function of rf power. Similar disagreement between the Bessel-function predictions and observations was mentioned in \cite{gallagher} in the case of resonant dipole-dipole collisional energy transfer of K Rydberg atoms, and was also attributed to AC Stark shift.

In conclusion, we have demonstrated a simple formula for the strength of coupling of colliding atoms to a molecular bound state when a magnetic field arbitrarily far from a Feshbach resonance is modulated in time. This formula is valid as long as the Zeeman shift is linear and the rf frequency is larger than the Feshbach width. We show that it is possible to efficiently associate molecules far from a Feshbach resonance, provided an rf Rabi frequency similar to the rf frequency is achievable. In our experiment, the detuning of the magnetic field from the Feshbach resonance was up to $10^5$ times the width of the Feshbach resonance. We have derived these results assuming that coupling is provided by dipolar interactions, and that there is no hyperfine structure. However, the model can readily be extended to any other coupling, and the results remain valid in presence of hyperfine structure, provided that the Zeeman effect is linear over a range of fields set by the amplitude of rf modulation.

Acknowledgements: LPL is Unit\'e Mixte (UMR 7538) of CNRS and of Universit\'e Paris Nord. We acknowledge financial support Minist\`{e}re de l'Enseignement Sup\'{e}rieur et de la Recherche (CPER), IFRAF (Institut Francilien de Recherche sur les Atomes Froids) and the Plan-Pluri Formation (PPF) devoted to the manipulation of cold atoms by powerful lasers.

\end{document}